\documentclass[twocolumn,showpacs]{revtex4}

\usepackage{graphicx}
\usepackage{dcolumn}
\usepackage{amsmath}

\makeatletter
\def\btt#1{\texttt{\@backslashchar#1}}%
\DeclareRobustCommand\bblash{\btt{\@backslashchar}}%
\makeatother

\begin{document}
\pagestyle{empty}
\preprint{MgAlB2 NMR}

\title{Evidence for High-frequency Phonon Mediated S-wave Superconductivity : 
$^{11}$B-NMR Study of Al-doped MgB$_2$}
\author{H.~Kotegawa$^{1}$, K.~Ishida$^{1}$, Y.~Kitaoka$^{1,3}$, T.~Muranaka$^{2}$, N.~Nakagawa$^{2}$, H.~Takagiwa$^{2}$, and J.~Akimitsu$^{2,3}$}

\address{$^1$Department of Physical Science, Graduate School of Engineering Science, Osaka University, Toyonaka, Osaka 560-8531, Japan\\
$^{2}$Department of Physics, Aoyama-Gakuin University, Setagaya-ku, Tokyo 157-8572, Japan.\\
$^{3}$Core Research for Evolutional Science and Technology (CREST) of the Japan Science and Technology Corporation (JST)}

\date{\today}

\begin{abstract}

We report $^{11}$B-NMR study on Al-doped MgB$_2$ that addresses a possible mechanism for a high superconducting (SC) transition temperature ($T_c$) of $\sim 40$ K in recently discovered MgB$_2$. The result of nuclear spin lattice relaxation rate $1/T_1$ in the SC state revealed that the size in the SC gap is not changed by substituting  Al for Mg. The reduction on $T_c$ by Al-doping is shown to be due to the decrease of $N(E_F)$. According to the McMillan equation, the experimental relation between $T_c$ and the relative change in $N(E_F)$ allowed us to estimate a characteristic phonon frequency $\omega \sim 700$ K and an electron-phonon coupling constant $\lambda \sim 0.87$. These results suggest that the high-$T_c$ superconductivity in MgB$_2$ is mediated by the strong electron-phonon coupling with high-frequency phonons.
\end{abstract}

\vspace*{5mm}

\maketitle

Recently discovered superconductor MgB$_2$, which exhibits a high value of superconducting (SC) transition temperature $T_c\sim 40$ K, has attracted enormous interests.\cite{Nagamatsu} Soon after its discovery, a lot of important experimental works were reported on isotope effect,\cite{Bud'ko} tunneling,\cite{Rubio,Karapetrov} photoemission,\cite{Takahashi} inelastic neutron,\cite{Osborn} and specific-heat measurements.\cite{Walti,Kremer}
 We reported  $^{11}$B-NMR measurements to investigate the SC and the normal state-properties. The NMR results unraveled that the SC gap is of an $s$-wave type with a magnitude of gap $2\Delta/k_BT_c \sim 5$, which suggested that it is in a strong coupling regime.\cite{Kotegawa} 
A lot of experiments including our NMR result suggest that the superconductivity of MgB$_2$ is well understood in terms of the BCS theory.  
On the basis of the band calculations,\cite{An,Kortus,Kong}
An and Pickett pointed out that holes in the $\sigma$ band at the Boron (B) layer are strongly coupled with the E$_{2g}$ phonon mode of $\sim 700$ K with a large value of electron-phonon coupling constant $\lambda \sim 1$.
In fact, this phonon mode has been observed by the Raman and inelastic-neutron scattering measurements.\cite{Osborn,Bohnen,Yildirim,Muranaka}
Yildirim {\it et al.} emphasized that the E$_{2g}$ mode, which is expected to be quite anharmonic, strongly couples with the electrons in the $\sigma$ band.\cite{Yildirim}
The direct experimental evidence is, however, not presented yet to show that the high-frequency E$_{2g}$ phonon mode is responsible for the onset of the high-$T_c$ superconductivity in MgB$_2$.

At present, the existence of two SC gaps has been accepted as suggested from several measurements \cite{Bouquet,Szabo,Chen,Tsuda,Giubileo,Yang} and from a theoretical model.\cite{Liu} 
Most experiments have, however, been made on either polycrystalline or its powdered form.
The band calculation shows that the Fermi surfaces consist of quasi-two dimensional (2D) cylindrical sheets and 3D-tubular networks, which seems to be consistent with the two SC gap model. 
Recently, a single crystal was synthesized, enabling us to obtain more detailed information on underlying SC characteristics.\cite{Lee,Xu} The on-going measurements have revealed an anisotropic SC characteristics in the upper critical field, $H_{c2}$. Furthermore, the Raman scattering measurement on the single crystal pointed to a single gap with $2\Delta/k_BT_c=3.9\pm 0.1$.\cite{Quilty} 
The presence of a small gap seems to depend on some quality of samples. 
Especially, the experiments investigating the surface such as tunneling, point contact, or photoemission measurement seem to show rather large ratio of the small gap to large one.
Further careful experiments and their analyses which give information about the bulk SC characteristics are highly desired.
In this meaning, NMR measurement is one of the best experiments since $1/T_1$ in the SC state, which is determined by the quasiparticles, would be affected significantly by the small gap.
Moreover, the NMR measurements are performed in the mixed state, then B nucleus within the radius of the penetration depth around the vortices can be detected, which reflect a bulk SC characteristics.

In order to gain further insight into the properties of the normal- and the SC state, and its SC mechanism, an impurity effect has been extensively investigated through the substitution for either Mg or B sites.  Slusky {\it et al.} reported that $T_c$ decreases by the Al substitution for Mg where electrons are doped into the $\sigma$ band at the Fermi level.\cite{Slusky}
Therefore, they argued that the decrease on $T_c$ is related to the reduction in  the density of states (DOS) at the Fermi level, $N(E_F)$ as suggested by the band calculation. They also revealed that a structural instability is induced when Al content $x$ exceeds 10\%.
In the range $0.1<x<0.25$, a non SC phase is segregated from a SC one and  
for $0.25<x$ where the lattice parameter along the $c$ axis becomes shorter than the undoped one, its bulk superconductivity is no longer observed. 
A moderate decrease of $T_c$ is considered to be intrinsic for the compounds  for $x\leq 0.1$.

It is known that nonmagnetic impurities do not affect the SC properties in the $s$-wave superconductors with an isotropic energy gap according to the theorem of dirty superconductors.\cite{Anderson} For an anisotropic $s$-wave case, any impurity scattering smears out the anisotropic gap over the Fermi surface, making it rather isotropic. As a result, this smearing effect in the gap structure appears in nuclear spin lattice relaxation ($1/T_1$) behavior.\cite{Mukuda} For example, the coherence peak in $1/T_1$ is more enhanced rather than the impurity undoped one. In the previous paper about non-doped MgB$_2$,\cite{Kotegawa} we reported that the $1/T_1$ in the SC state shows a tiny coherence peak just below $T_c$, followed by an exponential decrease below 0.8$T_c$. The systematic $T_1$ measurement on the Al-doped MgB$_2$ would provide valuable information about the causes for the significant suppression of the coherence peak and for the reduction in $T_c$.

In this paper, we argue that a two SC gaps model is not applicable in interpreting the NMR result on MgB$_2$, and report the results on (Mg$_{1-x}$Al$_x$)B$_2$ with $x=0.05, 0.07, 0.1$, and 0.2, focusing on the $^{11}$B-$T_1$ measurement. 
In the normal state, $1/T_1T$ in proportion to $N (E_F)^2$ decreases with increasing the Al content, demonstrating the reduction of  $N(E_F)$. 
In the SC state, the $1/T_1$ at $x=0.05$ resembles that in the undoped one, showing that the size of the gap remains unchanged.  
The reduction on $T_c$ is shown to be due to the decrease of $N(E_F)$. According to the McMillan equation, the experimental relation between $T_c$ and the relative change in $N(E_F)$ against the value of MgB$_2$ allows us to deduce a characteristic phonon frequency, $\omega \sim 700$ K and an electron-phonon coupling constant, $\lambda \sim 0.87$. These values suggest that a cause for the high $T_c$ value in MgB$_2$ might be due to the strong coupling between the electrons near the Fermi level and the high-frequency E$_{2g}$ phonon mode. 

Polycrystalline samples of Mg$_{1-x}$Al$_x$B$_2$ with $x=0$, 0.05, 0.07, 0.1, and 0.2 were prepared as in Ref.~\cite{Nagamatsu}, and used without powdering to avoid some crystal defect or oxidation if any. The x-ray diffraction revealed that the samples for $x\leq 0.1$ consist of a single phase, whereas the phase segregation occurs in the sample at $x=0.2$, which is in agreement with the result reported in ref.\cite{Slusky}. The value of $T_c$ was determined by the diamagnetization measurement using SQUID.

The $T_1$ measured at the central peak (the $1/2 \leftrightarrow -1/2$ transition) was determined by fitting the relaxation function of the nuclear magnetization $m(t)=(M(\infty)-M(t))/M(\infty)$, following the two-exponential form,\cite{Narath}
\begin{eqnarray}
 m(t) = \frac{1}{10}\exp\left(-\frac{t}{T_1}\right) +\frac{9}{10}\exp\left(-\frac{6t}{T_1}\right).
\end{eqnarray}
Here $M(t)$ is the nuclear magnetization at a time $t$ after saturation pulses.

In the normal state, a single component of $T_1$ was precisely determined, independent of the field. A $T_1T$ = constant relation was observed down to $T_c$ with a value of $T_1T=1.75\times 10^2$ ($s \cdot$ K).
In the SC state, however, the relaxation function $m(t)$ was not fitted with a single component of $1/T_1$ because of the anisotropy in $H_{c2}$ \cite{Lee,Xu} and the presence of the vortex core. Because of the anisotropy in $H_{c2}$, $T_c$ at each grain depends on its direction of $H$. NMR spectrum shows a typical powder pattern for nuclear spin $I=3/2$ due to quadrupolar effects, having peak for the fractions of $\theta = 90^{\circ}$ and $41.8^{\circ}$.\cite{Papavassiliou}
 Here, $\theta$ represents the angle between the magnetic field direction and $c$-axis.
Therefore, on average, the short and the long components in $T_1$ are expected to arise from the respective grains for $\theta \sim 41.8^{\circ}$  and $\sim 90^{\circ}$ ($H \perp c$) below $T_c(H_{\perp})=29$ K. 
In fact, the short component shows a similar behavior to the long component as if it indicates a $T_c \sim 19$ K, which seems to be $T_c(\theta \sim 41.8^{\circ})$.
Note that the short component includes also the other contributions such as relaxation associated with the vortex cores, or some non-SC fractions, if any. 
Thus, the long component of $T_1$ is expected to probe an intrinsic relaxation behavior in the SC state.

First of all, we claim that our NMR result is inconsistent with the exsistence of two SC gaps in MgB$_2$ reported in the literatures.\cite{Bouquet,Szabo,Chen,Tsuda,Giubileo,Yang}
Fig.~1 shows temperature ($T$) dependence of $(1/T_1T)_S/(1/T_1T)_N$ of $^{11}$B in MgB$_2$ that was reported previously and reexamined in the present work. Here $(1/T_1T)_S$ in the SC state is normalized by the value of $(1/T_1T)_N$ in the normal state.

By assuming two values of SC gap as $2\Delta_L/k_BT_c=5$ and $2\Delta_S/k_BT_c=1.6$, and taking a ratio of the respective fraction in the density of states $N(E_F)$ at the Fermi level as $N_L/N_S$=1, 3, and 5, a $T$ dependence of $1/T_1T$ 
is calculated as indicated by dashed lines in Fig.~1. The inset indicates the 
BCS-like $T$ dependence of two gaps below $T_c$.
Solid line indicates a result calculated by assuming an $s$-wave model with a single gap of $2\Delta/k_BT_c=5$.\cite{Kotegawa}
It is evident that any two-gaps model is not consistent with the NMR data. It should be noted that a value of $N_L/N_S<1$ was indicated by the other experiments.
The present NMR results, therefore, are against the presence of small gap, suggesting that it may be extrinsic for the superconductivity in MgB$_2$.
We should stress that the long component in $T_1$ is easily masked after either powdering sample or even safekeeping it in a vacuum for a few months, and 
eventually, $T_1$ shows a $T_1T={\rm constant}$ relation even far below $T_c$.
The presence of the small gap might arise from some aging effects, although many other experiments reconcile the presence of two SC gaps.

In Fig.~2 is shown the $T$ dependence of $1/T_1T$ for Mg$_{1-x}$Al$_x$B$_2$ with $x=0, 0.05, 0.07, 0.1$, and 0.2. 
First, we deal with the $1/T_1$ result in the SC state.  
The inset of Fig.~2 indicates $(1/T_1T)_S/(1/T_1T)_N$ vs $T/T_c(H)$ plots at $x=0.05$ along with the data on MgB$_2$.\cite{Kotegawa} 
A value of the tiny coherence peak $(1/T_1T)_S/(1/T_1T)_N\sim $1.2 was observed just below $T_c$ in MgB$_2$, and it remains unchanged even at $x=0.05$, followed by a behavior similar to $x=0$ upon cooling. It should be noted that the magnitude in SC gap is not reduced  by the Al substitution for Mg, suggesting the lack of pair-breaking effect.

Concerning the origin for the tiny coherence peak, we raise two possibilities 
of an anisotropic $s$-wave state and the damping effect of quasiparticles that  strongly interact with thermally excited phonons due to the high value of $T_c(H)$. In the former, the anisotropic SC gap may originate from the anisotropy of the Fermi surface and/or of the pairing interaction. If nonmagnetic impurities averaged over an anisotropy in the gap structure, the coherence peak in $1/T_1T$ could be even more enhanced rather than non-doped one.\cite{Anderson,Mukuda}
As seen in the inset of Fig.~2, the coherence peak at $x=0.05$ is almost the same as at $x=0$. Therefore, the impurity scattering does not have a significant effect on the tiny coherence peak. It should be noted that the coherence peak in NbB$_2$, which forms in the same crystal structure as MgB$_2$, is larger than in MgB$_2$ and its $T_c \sim 5$ K is much lower than $T_c=39$ K, showing small damping effect by the thermally excited phonons in NbB$_2$.\cite{Kotegawa2}
Therefore, the tiny coherence peak in $1/T_1$ on MgB$_2$ is considered as due to the damping effect of quasiparticles that strongly interact with thermally excited phonons and the strong coupling effect of electron-phonon interaction.

Why does $T_c$ decrease appreciably by the Al substitution for Mg ?
We show below that this is because the density of states $N(E_F)$ at the Fermi level is decreased with increasing the Al content.
It is evident from Fig.~2 that 
the magnitude in $(1/T_1T)_N$ decreases gradually with the Al content. 
In general, $1/T_1T$ is proportional to  $N(E_F)^2$ as follows:
\begin{eqnarray}
\frac{1}{T_1T} = \frac{\pi}{\hbar}A^2 N(E_F)^2 k_B,
\end{eqnarray}
where $A$ is the hyperfine coupling constant of $^{11}$B nuclear with conduction electrons. Accordingly, a relative change of $N(E_F)$ caused by the Al substitution is extracted from the relation of 
\begin{eqnarray}
\frac{N(E_F)}{N_0(E_F)}=\sqrt \frac{(1/T_1T)_x}{(1/T_1T)_{x=0}},
\end{eqnarray}
where $N_0(E_F)$ is the DOS of MgB$_2$.
The Al$^{3+}$ substitution for Mg$^{2+}$ introduces one electron into the $\sigma$ band at the Fermi energy. According to the band calculation, the decrease in $N(E_F)$ by the Al substitution is consistent with the Fermi surface characterized by holes. These results reveal that the suppression in $T_c$ originates from the decrease in $N(E_F)$.

The respective relative decreases in $T_c$ and $N(E_F)/N_0(E_F)$ are indicated in the inset of Fig.~3.
A linear decrease is seen on $T_c$ vs $x$ plots up to $x=0.2$, whereas $N(E_F)/N_0(E_F)$ deviates from a linear relation at $x=0.2$ where a non-SC phase is segregated from a SC one.\cite{Slusky}
The non-SC phase, which may presumably include the Al content higher than in the SC phase, might have a different $T_1$.
Therefore, we discuss on a possible relation between $T_c$ and $N(E_F)/N_0(E_F)$ up to $x=0.1$.
Fig.~3 indicates  $T_c$ vs $N(E_F)/N_0(E_F)$ plots with the Al content $x$ as the implicit parameter. 
Following the McMillan equation,\cite{McMillan,Allen}
\begin{eqnarray}
T_c=\frac{\omega}{1.2} \exp \left( \frac{-1.04(1+\lambda)}{\lambda-\mu^*(1+0.62\lambda)} \right),
\end{eqnarray}
, where $\omega$, $\lambda$ and $\mu^*$ are the phonon frequency, an electron-phonon coupling constant, and a Coulomb pseudpotential, respectively, and $\lambda(x)=N(E_F)V_{ph}$ and $\mu^*(x)=N(E_F)V_c$ are expressed using an electron-phonon interaction $V_{ph}$ and a Coulomb interaction $V_c$.
Now, we consider the obtained experimental relation on the basis of the above relation.
For simplicity, we may assume that $V_{ph}$, $V_c$, and $\omega$ are unchanged as long as Al content is in a small content.
As $N(E_F)$ changes, $\lambda$ and $\mu^*$ are modified following $\lambda=(N(E_F)/N_0(E_F))\cdot\lambda(x=0)$ and $\mu^*=(N(E_F)/N_0(E_F))\cdot\mu^*(x=0)$.
Using these relations and assuming a commonly-used value of $\mu^*(x=0)=0.10$, 
the experimental relation between $T_c$ and $N(E_F)/N_0(E_F)$ can be fitted by 
the McMillan equation with a different set of parameters like $\lambda(x=0)=0.8$, and $\omega=840$ K or $\lambda(x=0)=0.93$, and $\omega=630$ K. A best set of fitting parameters are deduced as $\lambda(x=0)=0.87\pm 0.06$, and $\omega=706\pm 96$ K. These values are in good agreement with the theoretical values.\cite{An,Kortus,Kong} Note that the experimental relation is never reproduced by assuming a Debye frequency of an acoustic phonon around $\omega=$ 300 K as shown in the figure.
A high phonon-frequency around 700 K is comparable to the frequency of the optical E$_{2g}$ mode obtained by the neutron-diffraction and the Raman-scattering experiments.\cite{Osborn,Bohnen,Yildirim,Muranaka} Above result strongly suggests that the strong coupling between the electrons near Fermi level in the $\sigma$ band and such the higher-frequency phonon as the E$_{2g}$ mode may play vital role for the onset of high-$T_c$ superconductivity in MgB$_2$.

In summary, the most NMR experiments are totally consistent with the strong coupling $s$-wave model with the single value of the large gap of 2$\Delta/k_BT_c=5$.
The $^{11}$B-NMR measurements on Mg$_{1-x}$Al$_x$B$_2$ revealed that the Al substitution for Mg does not bring about the pair breaking.
The reduction in $N(E_F)$ has been clearly evidenced from the reduction in $1/T_1T$ in proportion to $N(E_F)^2$. These results demonstrated that the reduction in $N(E_F)$ with the Al content is responsible for the decrease in $T_c$. According to the McMillan equation, the experimental relation between $T_c$ and the relative change in $N(E_F)$ allowed us to estimate a phonon frequency, $\omega\sim 700$ K and an electron-phonon coupling constant $\lambda\sim 0.87$. These results are in good agreement with the theoretical suggestion that the high $T_c$ superconductivity in MgB$_2$ is mediated by the strong electron-phonon coupling with high frequency-phonons.

\begin{figure}[htbp]
\begin{center}
\includegraphics[width=\linewidth]{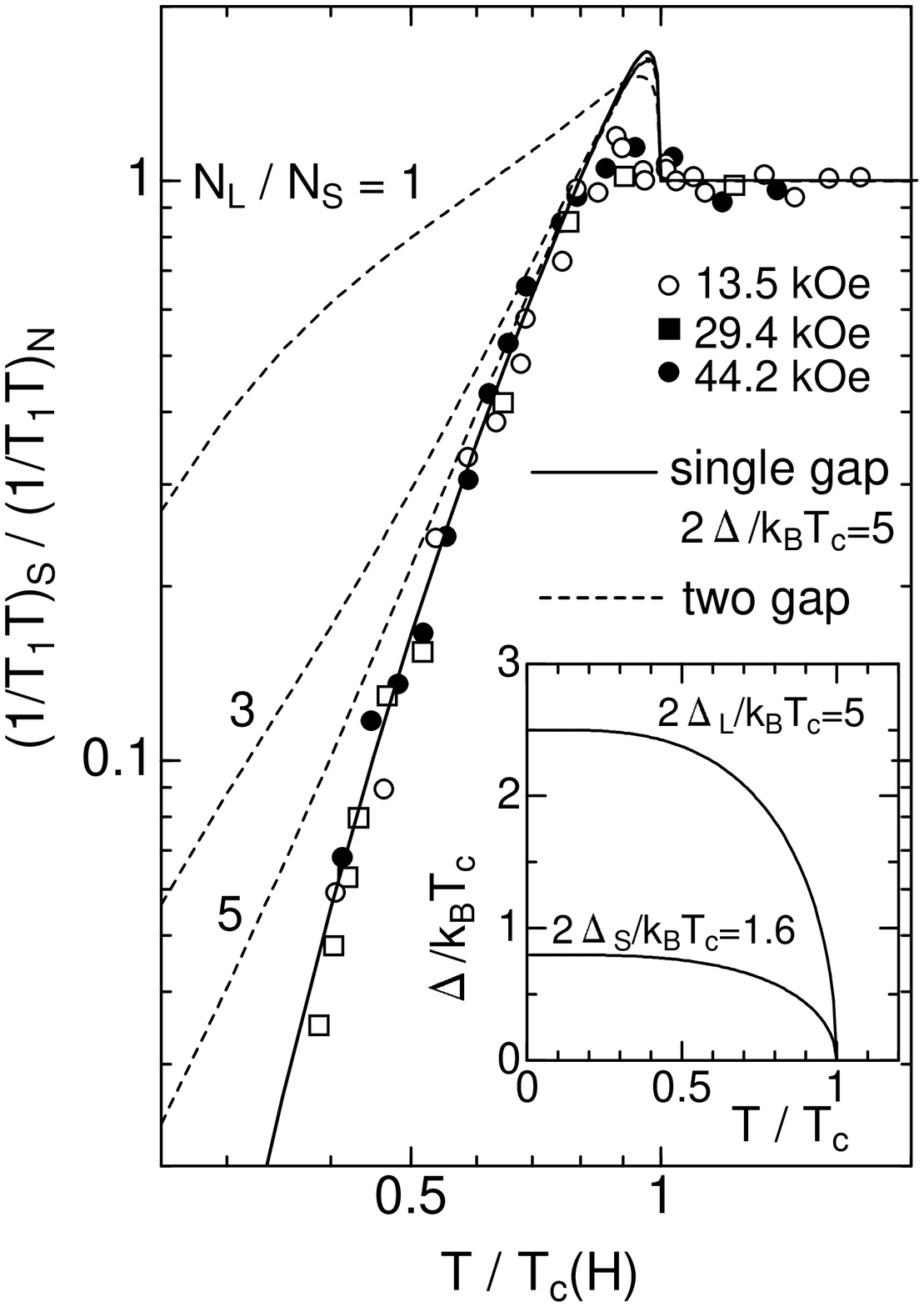}
\caption[]{\protect $T$ dependencies of $(1/T_1T)_S/(1/T_1T)_N$ of $^{11}$B in MgB$_2$. Solid line is a calculation in terms of an $s$-wave model with a single isotropic gap of 2$\Delta/k_BT_c=5$. Dashed lines are calculations on a two-gap model with 2$\Delta_L/k_BT_c=5$ and 2$\Delta_S/k_BT_c=1.6$, assuming the BCS-like $T$ dependence below $T_c$ as indicated in the inset. 
Here, a ratio of the fraction in the density of states at the Fermi level is used as $N_L/N_S=$ 1, 3 and 5.}
\end{center}
\end{figure}

\begin{figure}[htbp]
\begin{center}
\includegraphics[width=\linewidth]{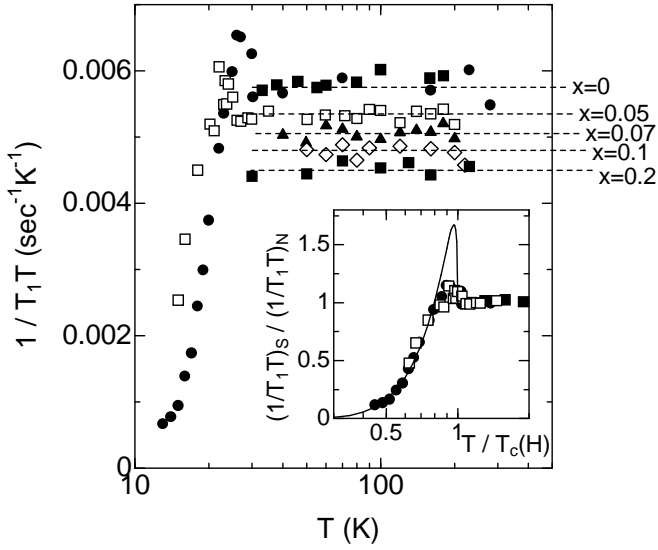}
\caption[]{\protect $T$ dependence of $^{11}(1/T_1T)$ in Mg$_{1-x}$Al$_x$B$_2$ ($x=0, 0.05, 0.07, 0.1$ and 0.2) at the magnetic field $H=4.42$ T.
The data in  MgB$_2$ at $H=1.35$ T (4.42 T) are shown by squares (circles). The inset indicates a comparison of the behavior in $(1/T_1T)$ below $T_c$ at $x=0$ and 0.05, revealing nearly the same behavior.
}
\end{center}
\end{figure}

\begin{figure}[htbp]
\begin{center}
\includegraphics[width=\linewidth]{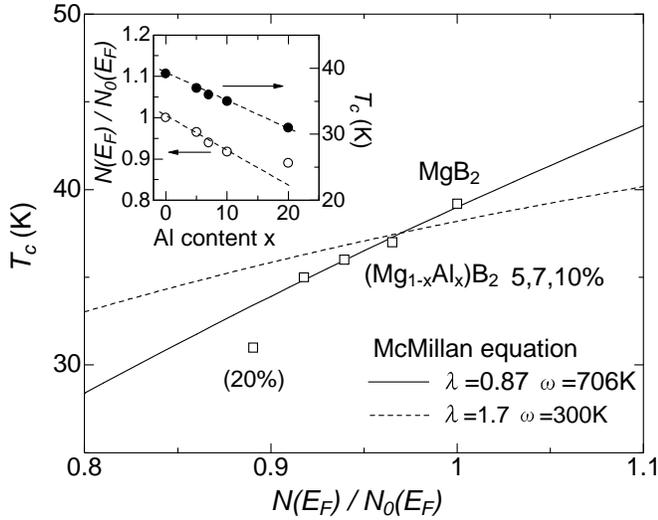}
\caption[]{\protect The dependence of $T_c$ on $N(E_F)/N_0(E_F)$.
Solid line corresponds to a best fitting with a phonon frequency $\sim 700$ K and a electron-phonon coupling constant, $\lambda\sim 0.87$ on the basis of the McMillan equation (see text).
The dashed curve is obtained by assuming an acoustic Debye frequency with 300 K.
Here the datum at $x=0.2$ is omitted because of its phase segregation.
The inset is the dependencies of $T_c$ and $N(E_F)/N_0(E_F)$ on the Al-content ($x$).
}
\end{center}
\end{figure}

\end{document}